# Evidence for High $T_c$ Superconducting Transitions in Isolated $Al_{45}^-$ and $Al_{47}^-$ Nanoclusters

Baopeng Cao,[1] Colleen M. Neal,[1] Anne K. Starace,[1] Yurii N. Ovchinnikov,[2] Vladimir Z. Kresin,[3] and Martin F. Jarrold[1]

[1] Chemistry Department, Indiana University, 800 E. Kirkwood Ave., Bloomington, IN 47405, USA

[2] L. Landau Institute for Theoretical Physics, 117334, Moscow, Russia

[3] Lawrence Berkeley Laboratory, University of California, Berkeley, California 94720, USA

**Abstract**

**Heat capacities measured for $Al_{45}^-$ and $Al_{47}^-$ nanoclusters have reproducible jumps at ~ 200 K. These jumps are consistent with theoretical predictions that some clusters with highly degenerate electronic states near the Fermi level will undergo a transition into a high $T_c$ superconducting state. An analysis based on a theoretical treatment of pairing in $Al_{45}^-$ and $Al_{47}^-$ agrees well with the experimental data in both the value of the critical temperature and in the size and width of the jumps in the heat capacity.**

Superconductivity, the flow of electric current without resistance, occurs in a wide variety of materials when the temperature drops below a critical temperature, $T_c$. The transition to the superconducting state is accompanied by diamagnetism and by a jump in the heat capacity. The superconducting transitions observed to date are all below 135 K under atmospheric pressure and below 168 K under high pressure, and they are mostly associated with bulk systems [1-3]. Recently, Kresin and Ovchinnikov proposed that a superconducting transition can occur for some metal nanoclusters with $10^2 - 10^3$ valence electrons [4-5]. The transitions are expected to occur near electronic shell closings. It is well-established that the electronic energy levels in metal clusters are not evenly spaced, but instead adopt a shell structure similar to that found in atoms or nuclei [6-10]. The highest energy occupied electronic energy levels become highly degenerate at spherical shell closings ("magic" numbers) while between shell closings Jahn-Teller distortions lift the degeneracy. Under special circumstances, which we describe below, the shell structure is expected to strengthen pairing correlations, and lead to elevated values for the critical temperature for a transition to the superconducting state. Superconducting state of small nanoclusters is directly related to the phenomenon of pair correlation. For bulk superconductors it leads to an appearance of the dissipationless state. For the clusters of interest the manifestation is different and is similar to that in atomic nuclei: the pairing leads to a strong modification of the energy spectrum.

In this letter we report heat capacity measurements for isolated, single-sized aluminum nanoclusters. The heat capacity measurements were performed for a number of aluminum nanoclusters, and both $Al_{45}^-$ and $Al_{47}^-$ were found to display a clear, reproducible jump in their heat capacities at around 200 K. As we show below, this behavior is consistent with theoretical predictions for a superconducting transition in aluminum nanoclusters [4-5]. A transition temperature of 200 K exceeds those found in bulk systems. While it is difficult to imagine direct applications for superconductivity in isolated nanoclusters, these results suggest the feasibility of generating superconducting nanowires by aligning nanoclusters on a surface or in a nanotube. The nature of the nanocluster (i.e. number of atoms and perhaps even the composition) could then be adjusted to optimize the electronic structure for high $T_c$ superconductivity. The specific alignment of size-selected nanoclusters on a surface or in a nanotube is beyond our current capabilities; however, the technological hurdles are not insurmountable.

The heat capacities were measured using a novel multi-collision induced dissociation method where the dissociation threshold is determined as a function of the cluster's initial temperature [11-14]. As the temperature is raised, the dissociation threshold decreases due to the increase in the cluster's internal energy. The change in the dissociation threshold with temperature provides a measure of the heat capacity. The apparatus used to measure the heat capacities has been described in detail elsewhere [11-14]. Briefly, aluminum cluster anions are generated by pulsed laser vaporization of a liquid aluminum target [15]. Once formed, the cluster ions travel into a temperature variable extension where their temperature is set. Upon exiting the extension, the cluster ions are focused into a quadrupole mass spectrometer where a particular cluster size is selected. The size-selected clusters are then focused into a high pressure collision cell, where they are heated by collisions with the He buffer gas and may dissociate. The undissociated parent ions and the fragment ions are subsequently analyzed by a second quadrupole mass spectrometer. The fraction of the cluster ions that dissociate is determined from the mass spectrum as a function of the ion's translational energy, and the translational energy required for 50% of the cluster ions to dissociate (TE50%D) is determined from a linear regression. The derivative of TE50%D with respect to temperature is proportional to the heat capacity. The proportionality constant is related to the fraction of the ion's translational energy that is converted into internal energy as the ion enters the collision cell. This fraction is estimated from a modified impulsive collision model [16-18]. Figure 1 shows plots of heat capacities determined for aluminum cluster anions with 43-48 atoms for temperatures below room temperature. The heat capacities are plotted relative to the classical value $(3n - 6 + 3/2)k_B$, where n is the number of atoms in the cluster. The thin dashed lines in the figure show heat capacities derived from a modified Debye model that takes into account the finite size of the cluster [19]. There are peaks in the heat capacities for all of the clusters at temperatures substantially above 300 K. These peaks are due to melting transitions, and in some cases pre-melting transitions, and are discussed in more detail elsewhere [20]. In the heat capacities recorded below 300 K (see Figure 1) there are reproducible jumps at around 200 K for both $Al_{45}^-$ and $Al_{47}^-$. These jumps are are not due to melting or premelting transitions because they occur at too low a temperature. It is unlikely that they are due to structural transitions for the same reason. The structural transitions we have observed with aluminum clusters lead to dips rather than peaks in the heat capacities, and they occur at temperatures significantly above room temperature [12].

As noted above and demonstrated below, we believe that the observed jumps in heat capacity at 200 K for $Al_{45}^-$ and $Al_{47}^-$ correspond to the transition into a superconducting state. A superconducting transition has been observed at low temperature for large nanoparticles (N $\gtrsim 10^4$, N is the number of delocalized electrons) [21-24] and a nascent superconducting state has been invoked to explain some unexpected observations for small niobium clusters [25].

Al clusters are of special interest, because they display an electronic shell structure [26-28]. It has been suggested that the presence of an electronic shell structure can have a dramatic impact on electron pairing [29,30]. Note that atomic nuclei (see[31 ], [32 ]) ,nanoparticles and granular metals (see,e.g.,[33 ]) are the examples of finite systems ; one can observe various manifestations of the superconducting state (Cooper pairing) in these systems.For nanoclusters one can observe the odd-even effect in their spectra, diamagnetism, jump in heat capacity, Josephson tunneling for the cluster network, etc.

According to [4,5] ,for clusters with parameters satisfying special but realistic conditions one can expect a great strengthening of the pairing correlation and, correspondingly, a large increase in the critical temperature [4,5]. These conditions are: the proximity of the electronic state to a complete shell ("magic" number) and a relatively small gap between the highest occupied shell (HOS) and the lowest unoccupied shell (LUS). $Al_{45}^-$ and $Al_{47}^-$ both satisfy these criteria. The electronic shell structure for aluminum clusters as well as for gallium clusters (gallium and aluminum atoms have a similar electronic structure) has been observed experimentally [26-28]. According to photoionization data [26-28], both $Al_{45}^-$ and $Al_{47}^-$ are close to a shell closing. A high $T_c$ is caused by large values of orbital momenta $l$ and corresponding large shell degeneracy [4,5]. That is why the situation is most favorable for relatively large clusters with N ≳ $10^2$. According to theoretical predictions, the values of $T_c$ observed in the present paper are realistic. Indeed, consider for example the cluster $Al_{45}^-$. This cluster contains an even number of electrons (this is essential for the pair correlation) and it is close to the shell closing with N=138 [6-10, 26-28]. The equation for the pairing order parameter $\Delta(\omega_n)$ has the following form [4,5] :

$$\Delta(\omega_n) Z = \eta \frac{T}{2V} \sum_{\omega_{n'}} \sum_s D(\omega_n - \omega_{n'}) F_s^+(\omega_{n'}) \qquad (1)$$

Here

$$D(\omega_n - \omega_{n'}, \tilde{\Omega}) = \tilde{\Omega}^2 \left[ (\omega_n - \omega_{n'})^2 + \tilde{\Omega}^2 \right]^{-1} \quad ; \quad F_s^+(\omega_{n'}) = \Delta(\omega_{n'}) \left[ \omega_{n'}^2 + \xi_s^2 + \Delta^2(\omega_{n'}) \right]^{-1} \qquad (1')$$

are the vibrational propagator and the Gor'kov pairing function (see, e.g. [34]) respectively, $\omega_n = (2n+1)\pi T$ ; n = 0, ± 1, ± 2,… , $\xi_s = E_s - \mu$ is the energy of the s'th electronic state referred to the chemical potential $\mu$ , V is the cluster volume, $\eta = <I>^2 / M\tilde{\Omega}^2$ is the Hopfield parameter, <I> is the electron-ion matrix element averaged over electronic states involved in the pairing M is the ionic mass, and Z is the renormalization function. Equation (1) looks similar to that in the theory of strong coupling superconductivity (see, e.g. [35-37]) , but is different in two key aspects. Firstly, it contains a summation over discrete energy levels $E_S$ and, in addition, the position of the chemical potential is determined by the values of N and T. Note that in a weak coupling case ($\eta/V \ll 1$ and correspondingly $\pi T_C \ll \tilde{\Omega}$), one should put in Eq. (1) Z=1, D=1, recovering the usual BCS scheme (see,e.g.,[36],[37]). For a "magic" cluster, $T_c$ can be calculated from the Eq.(1) which at T=$T_c$ can be reduced to the following matrix equation (see Ref. [4,5]):

$$det|1 - K_{nn'}| = 0 \qquad (2)$$

where,

$$K_{nn'} = g\tau_c \sum_j G_j \{[1+(n-n')^2 \tau_c^2]^{-1} - \delta_{nn'} Z_n\} [(n'+0.5)^2 \tau_c^2 + \tilde{\xi}_j^2]^{-1} \qquad (2')$$

Here $\tau_c = 2\pi T_c/\tilde{\Omega}$, $G_j = 2(2l_j + 1)$, $g = \lambda_b(6\pi \tilde{E}_F/N)$, $\tilde{E}_F = E_F \tilde{\Omega}^{-1}$, $\Delta \tilde{E} = \Delta E \tilde{\Omega}^{-1}$, $\tilde{\xi}_j = \xi_j \tilde{\Omega}^{-1}$, $j \equiv \{H,L\}, H \equiv HOS, L \equiv LUS$, $\tilde{\Omega}$ is the characteristic vibrational frequency of the cluster, $\tilde{\Omega} \approx \Omega_D$, $l$ is the orbital quantum number, $E_F$ is the Fermi energy, and $Z_n$ is the renormalization function. The matrix equation (2) converges rapidly for $\tau_c > 1$.

For $Al_{45}^-$ clusters, $l_{HOS}$ = 1 and $l_{LUS}$ = 7, so that $G_H + G_L$ = 36. According to measurements, $\Delta E \approx$ 40 meV [24]. Substituting the parameters N = 136, $E_F \approx$ 1.3×10$^5$ K, $\tilde{\Omega} \approx 350 K$ and $\lambda_b \approx$ 0.4 [4,5] into Equations (1) and (2), we obtain $T_c \approx$ 170 K. Jump in the heat capacities was experimentally observed at temperatures close to this value (see Figure 1). Note that high value of $T_c$ leads the coherence length being short, similarly to the high $T_c$ cuprates.

One can also evaluate the size of the jump in heat capacity. Let us start with the Ginzburg-Landau expression for the thermodynamic potential $\Theta_s$ in the superconducting state near $T_c$:

$$\Theta_s = \Theta_n - a\tau_c^2(1-t)^2 \qquad (3)$$

Here, $\Theta_n$ is the thermodynamic potential in the normal state, $t = T/T_c$, and a is the material dependent constant introduced in [4,5] and determined by the relation:

$$a = \tau_c^2 \tilde{S}/\pi(\partial \tau_c/\partial g)C \qquad (4)$$

where

$$\tilde{S} = 0.5 \sum_{\substack{s,s' \\ n,n' \geq 0}} (f^+_{n;n+1} + f^-_{n,n'})\chi_{n;\xi_s}\chi_{n';\xi_s'}\phi_n\phi_{n'} \qquad (5)$$

$$f^{\pm}_{n;r} = [1 + (n \pm r)^2\tau_c^2]^{-1} \qquad \chi_{n;v} = [(n+0.5)^2\tau_c^2 + v^2]^{-1} \qquad (6)$$

C determines the energy gap near $T_c$, g is the electron-vibrational coupling constant, and $\phi_n$ is the order parameter. A detailed calculation of the parameter a for the $Al_{45}^-$ cluster will be described elsewhere. However, for many clusters its value is of order of $a = (3 - 5) \times 10^3$ K. With use of Eq. (4) through (6), one can obtain the following expression for the jump in heat capacity:

$$\beta \equiv (C_s - C_n)/3n = (4\pi^2/3)(a/T_c)(T_c/\tilde{\Omega})^2 n^{-1} \qquad (7)$$

We normalized the jump to its classical value 3n (we set $k_B = 1$). Substituting into (7) the values of a, $T_c$, n and $\tilde{\Omega} \approx 30$ meV, we obtain $\beta \approx 0.3 \sim 0.5$ for the clusters of interest, in good agreement with the jumps found experimentally (see Figure ).

The observed jumps are rather broad. The breadth of the transition is caused by thermodynamic fluctuations. This question has been considered previously [4,5]. The thermodynamic fluctuations lead to

$$(\delta T_c/T_c) \approx (\tilde{\Omega}^2/\pi^2 a T_c)^2 \qquad (8)$$

Substituting the values for the parameters from above into this expression, we obtain ($\delta T_c/T_c$) ≈ 10 ~ 15%, that is, $\delta T_c \approx$ 30 K. This width corresponds quite well to the breadth of the transition observed for the $Al_{45}^-$ and $Al_{47}^-$ clusters. Therefore, the jumps observed in heat capacity for size selected aluminum clusters are entirely consistent with a high $T_c$ superconducting pairing transition predicted by theory [4,5].

In summary, heat capacities measured for size-selected $Al_{45}^-$ and $Al_{47}^-$ show jumps at around 200 K. The jump in the heat capacity is the signature of a phase transition. The features of the phase transition, i.e. the transition temperature, the size and the width of the peak in the heat capacity are all entirely consistent with a recent theoretical treatment of a superconducting transition for metal nanoclusters. This is the first experimental observation of such a transition.

**Acknowledgements:**

We are grateful to J. Friedel, L. Gor'kov, and V. V. Kresin for interesting discussions. The research of VK was supported by DARPA. MFJ thanks the NSF for financial support.

**References**.

**Figure Caption:**

Figure 1.   Heat capacities for the size-selected, aluminum nanoclusters with 43-48 atoms recorded below 300 K. The points are the results of multiple independent measurements, and the blue vertical bars show uncertainties (± one standard deviation). The heat capacities are plotted relative to the classical value $(3n - 6 + 3/2)k_B$, where n is the number of atoms in the cluster. The thin dashed lines show heat capacities derived from a modified Debye model that takes into account the finite size of the cluster. The obvious jumps observed at around 200 K in the heat capacity for $Al_{45}^-$ and $Al_{47}^-$ nanoclusters are believed to be due to the superconducting transition.

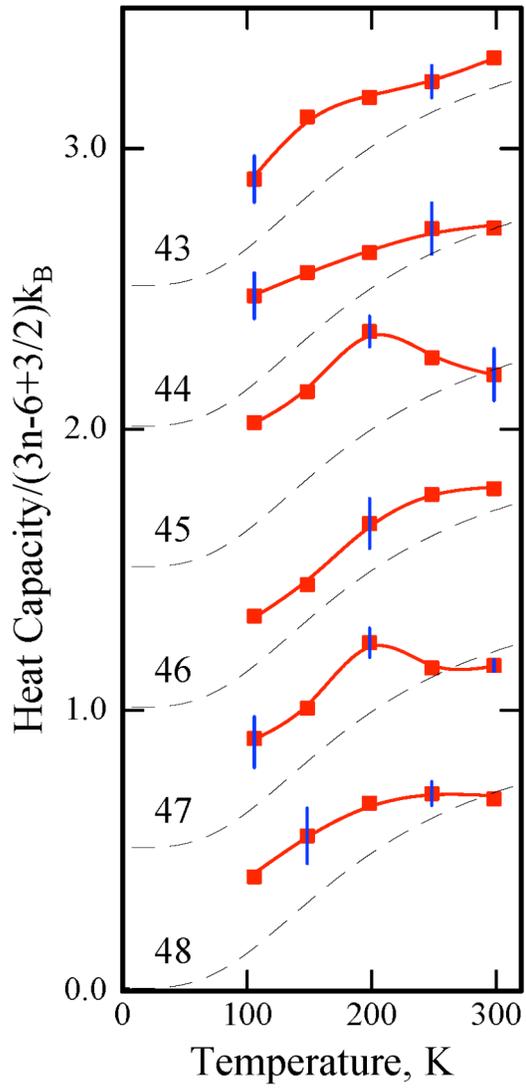